\documentclass[aps,prb,reprint,superscriptaddress,showpacs]{revtex4-1}
\usepackage{graphicx}
\usepackage{color}
\usepackage{transparent}
\usepackage{bm}
\usepackage{hyperref}
\usepackage{natbib}
\usepackage[export]{adjustbox}
\usepackage{sidecap}

\begin{document}
\title{Resonant optical control of the structural distortions that drive ultrafast demagnetization in Cr$_2$O$_3$}
\author{V. G. Sala}
\affiliation{ICFO-Institut de Ci\`{e}ncies Fot\`{o}niques, The Barcelona Institute of Science and Technology, 08860 Castelldefels (Barcelona), Spain}
\author{S. Dal Conte}
\affiliation{IFN-CNR, Dipartimento di Fisica, Politecnico di Milano, Milano, Italy}
\author{T. A. Miller}
\affiliation{ICFO-Institut de Ci\`{e}ncies Fot\`{o}niques, The Barcelona Institute of Science and Technology, 08860 Castelldefels (Barcelona), Spain}
\author{D. Viola}
\affiliation{IFN-CNR, Dipartimento di Fisica, Politecnico di Milano, Milano, Italy}
\author{E. Luppi}
\affiliation{Universit\'{e} Pierre et Marie Curie and CNRS, Paris, France}
\author{V. V\'{e}niard}
\affiliation{Laboratoire des Solides Irradi\'{e}s, \'{E}cole Polytechnique, Route de Saclay, F-91128 Palaiseau and European  Theoretical Spectroscopy Facility (ETSF), France}
\author{G. Cerullo}
\affiliation{IFN-CNR, Dipartimento di Fisica, Politecnico di Milano, Milano, Italy}
\author{S. Wall}
\email[Corresponding author:]{simon.wall@icfo.es}
\affiliation{ICFO-Institut de Ci\`{e}ncies Fot\`{o}niques, The Barcelona Institute of Science and Technology, 08860 Castelldefels (Barcelona), Spain}

\date{\today}

\begin{abstract}
We study how the color and polarization of ultrashort pulses of visible light can be used to control the demagnetization processes of the antiferromagnetic insulator Cr$_2$O$_3$. We utilize time-resolved second harmonic generation (SHG) to probe how changes in the magnetic and structural state evolve in time. We show that, varying the pump photon-energy to excite either localized transitions within the Cr or charge transfer states, leads to markedly different dynamics. Through a full polarization analysis of the SHG signal, symmetry considerations and density functional theory calculations, we show that, in the non-equilibrium state, SHG is sensitive to {\em both} lattice displacements and changes to the magnetic order, which allows us to conclude that different excited states couple to phonon modes of different symmetries. Furthermore, the spin-scattering rate depends on the induced distortion, enabling us to control the timescale for the demagnetization process. Our results suggest that selective photoexcitation of antiferromagnetic insulators allows fast and efficient manipulation of their magnetic state. 
\end{abstract}

\maketitle

\section{Introduction}
Selective control with resonant light pulses is an emerging route for manipulating the properties of transition metal oxides. Selectivity has been achieved through resonant excitation of magnetic dipole modes in antiferromagnets~\cite{Kampfrath2011} and multiferroics~\cite{Kubacka2014}, or through IR-active phonon modes, in order to drive insulator-metal transitions~\cite{Rini2007} and superconductivity~\cite{Fausti2011, Mitrano2016}.   

Ultrafast control of the magnetic state, in particular, could have a strong impact on magnetic recording technology. Ultrafast demagnetization is a complex process involving strong coupling between electronic, spin, and structural degrees of freedom which is dependent on the type of magnetic order and band structure~\cite{Koopmans2010,Kirilyuk2010}. Controlling these interactions is key for developing magnetic devices that can fully exploit femto-magnetism. Antiferromagnetic (AFM) materials are insensitive to external magnetic fields~\cite{Marti2014a} and are more stable when miniaturized ~\cite{Loth2012} than ferromagnets. Moreover the absence of a net magnetic moment can enable much faster control of spin dynamics ~\cite{Kimel2004}. 

Despite these possibilities, spin dynamics in AFM materials are still not well understood. Conventional techniques used to study the femtosecond dynamics of magnetic materials cannot be applied to AFM materials as they lack a net magnetic moment and alternative techniques are required. The advent of free electron lasers has enabled the study of spin dynamics through resonant elastic ~\cite{Ehrke2011a, Forst2011, Tobey2012, Kubacka2014} or inelastic magnetic scattering~\cite{Dean2016}. However, optical techniques such as linear magnetic birefringence~\cite{Bossini2014, Bossini2015} or second harmonic generation (SHG)~\cite{Fiebig2005} can also be used. 

Due to the small band gaps in many transition metal oxides, selective control has focused on the THz or mid-IR spectral region, which may reduce the speed at which the material can be driven. This is observed in the manganites, where it has been shown that the melting of antiferromagnetism via IR-active phonons is an order of magnitude slower and less efficient than when exciting above the band gap at 1.5\,eV~\cite{Forst2011}.

Cr$_2$O$_3$ is an ideal material to demonstrate ultrafast control of demagnetization at higher speeds. The band-gap is in the near UV, enabling optical control over a broader range of wavelengths. The magnetic state has been well characterized by SHG spectroscopy~\cite{Fiebig1994, Fiebig2005}, thus experiments can be performed with a regular laboratory setup and the static optical and electronic properties of the material are significantly better understood than those of other correlated materials, enabling an greater chance for understanding the dynamics induced by the laser pulse.

In this paper, we show that electronic visible photoexcitation can be used to selectively control the structural and magnetic properties of Cr$_2$O$_3$. By measuring the time- and polarization-resolved second harmonic signal, we show that different electronic in-gap states couple to different phonon modes. As these modes are responsible for the spin scattering process, we can control the demagnetization rate by as much as 25\% by changing the photon energy used to excite the system.

\section{Sample Characterization and Experimental Details}

\begin{figure}
\centering
\includegraphics{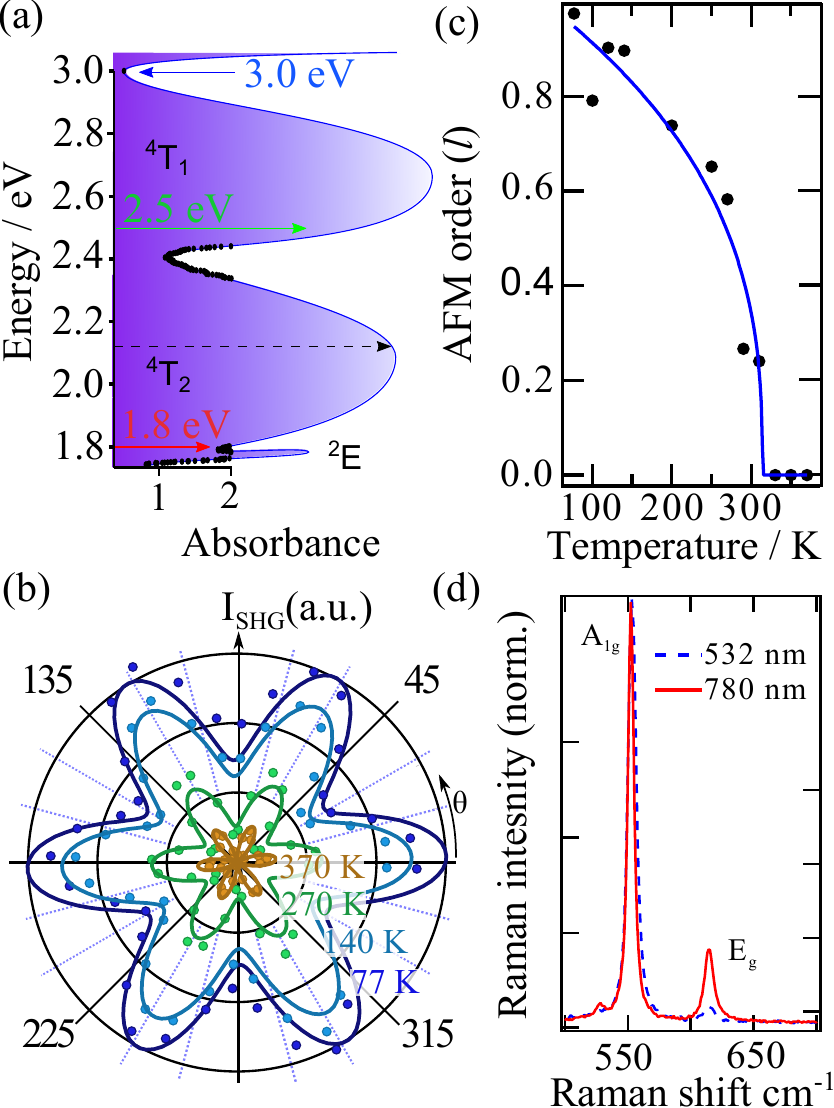}
\caption{(a) The absorbance of Cr$_2$O$_3$. Black markers correspond to measured values of our 10 $\mu$m thick sample at 77\,K. Blue line is a guide to the eye based on the measurements found in Ref~\cite{McClure1963}. The $^4T_1$ and  $^4T_2$ states correspond to transitions within crystal field split states of the Cr 3$d$ levels which preserve the spin of the electron. The $^2E$ state corresponds to a spin-forbidden transition. The colored arrows indicate the pump wavelengths used in the time resolved experiments and the black dashed arrow corresponds to the second harmonic of the probe. (b) Polarization and temperature dependence of the SHG signal from Cr$_2$O$_3$ measured with a fundamental photon energy of 1.08\,eV. Solid lines are fits to the data using Eq.~\ref{eq:angle}. (c) Temperature dependence of the antiferromagnetic order parameter $l(T)$ extracted from the fits from (b) giving $T_\mathrm{N} = 314\pm5$\,K. (d) Raman spectrum for two different excitation wavelengths, demonstrating coupling to $E_g$ modes is stronger for longer wavelengths at room temperature. 
\label{fig:Static}}
\end{figure}

A single crystal of Cr$_2$O$_3$, cut perpendicular to the trigonal $z$ axis, was grown using the flux method and was polished to a thickness of approximately 10\,$\mu$m and, unless otherwise stated, cooled to 77\,K. The measured sample absorbance is shown in Fig~\ref{fig:Static}a. Due to the small sample size and thickness it was not possible to measure attenuations greater than 2 orders of magnitude due to the dominance of scattered light around the sample edges. Optially, Cr$_2$O$_3$ can be considered a 100\% doped ruby crystal. The absorbance shows two broad peaks ($^4T_1$ and $^4T_2$), corresponding to crystal field excitations within the 3$d$ levels of the Cr ion which preserve the spin orientation of the photo-excited electron. The resonances are broadened due to a strong electron phonon coupling. In addition, a low energy narrow resonance ($^2E$) is observed which corresponds to a dipole forbidden transition in which the spin is flipped but the spatial part of the wavefunction is the same as the ground state~\cite{Muto1998}. The high contrast between the transmission of light on and between the resonances and the narrow Raman spectra (Fig~\ref{fig:Static}d) indicate that the sample is of high quality.  In the ruby laser, the $^4T_1$ and $^4T_2$ levels are optically pumped and the excited electrons decay non-radiatively to the $^2E$ state on sub picosecond timescales~\cite{Fonger1975}. As this process results in a spin flip, the lifetime of the excited state is very long and is used for lasing. Cr$_2$O$_3$, unlike Ruby, is anti-ferromagnetic, so the spin flip process can lead to rapid demagnetization. In this paper we track this process using time-resolved SHG. 

Cr$_2$O$_3$ is centrosymmetric with the crystallographic point group $\bar{3}m$. The presence of inversion symmetry forbids conventional SHG of the electric field, $E$; however an axial source term $\bm M \propto \bm{\chi_m^{(2)}}\bm{EE}$ is allowed. Below the N\'eel temperature, $T_\mathrm{N}= 308$\,K, the Cr ions order antiferromagnetically along the trigonal z-axis, enabling an additional polar source term $\bm P \propto\bm{\chi_e^{(2)}}\bm{EE}$. Both $\bm{\chi_e^{(2)}}$ and $\bm{\chi_m^{(2)}}$ satisfy the symmetry relations of the point group 32: $\chi_{xxx}^{(2)}=-\chi_{xyy}^{(2)}=-\chi_{yxy}^{(2)}=-\chi_{e/m}^{(2)}$~\cite{Boyd2003}. For light traveling along the trigonal $z$ axis, this gives rise to a SHG scattering pattern that depends on the linear polarization in the $(x, y)$ plane as
\begin{equation}
I_{\mathrm{SHG}}(\theta) \propto |\chi_e^{(2)}\sin(3\theta)-\chi_m^{(2)}\cos(3\theta) |^2 I_\mathrm{f}^2,
\label{eq:angle}
\end{equation}
where $I_\mathrm{f}$ is the intensity of the fundamental beam and $\theta$ is the polarization angle relative to the crystallographic $y$-axis. Furthermore, the non-linear susceptibilities can be expanded as a function of the antiferromagnetic order parameter, $l$, as $\chi_e^{(2)} = c_1 l(T)$ and $\chi_m^{(2)} = c_2 + c_3 l^2(T)$~\cite{Muto1998, Tanabe1998, Muthukumar1996}. 

The $c_i$ coefficients are temperature independent, but depend on the photon energy of the light used to probe the sample; resonant enhancement occurs for photon energies around 1.08\,eV due to the crystal field effects in the final state~\cite{Tanabe1998, Fiebig1994}. Thus resonant SHG is both sensitive to the magnetic order and the crystallographic structure surrounding the Cr$^{3+}$ ions. 

The temperature dependence of the static SHG signal was measured with the output of an optical parametric amplifier (OPA), pumped by the output of a Ti:sapphire laser at a repetition rate of 5 kHz and tuned to the 1.08 eV (1150\,nm) resonance. The pulse duration was approximately 60 fs. The SHG signal was measured in transmission with a high and low pass filter to separate out the 2.16\, eV (575\,nm) SHG signal from the fundamental and third harmonic light. A polarizer and half-waveplate set the linear polarization of the probe beam before the sample and a second polarizer after the sample selects the SHG signal at a polarization parallel to the incident beam. These were rotated to measure the polarization dependence of the SHG for the different temperatures shown in Fig.~\ref{fig:Static}b. From this data, the complex $c_i$ coefficients and the temperature-dependent order parameter, $l$, (Figs.~\ref{fig:Static}c) were determined, in good agreement with the literature~\cite{Fiebig1994, Fiebig2005}. 

For the time resolved experiments, a second OPA generated the pump pulse with central wavelengths at 3\,eV (400\,nm),  2.5\,eV (500\,nm), and 1.8\,eV (700\,nm) as indicated in Fig.~\ref{fig:Static}a, with pulse durations of order 50\,fs. These pump wavelengths are chosen in order to populate different excited states. 2.5\,eV and 1.8\,eV photons predominately populate the $d$ levels of the Cr ion, while the higher energy 3\,eV excitation also causes charge transfer excitation between O and Cr ions. Furthermore, the photon energies are detuned from the peak absorption to ensure a greater penetration depth of the pump light. 

The pump is set in a counter-propagating configuration, relative to the probe, exciting the sample from the backside and at a small angle as shown in the insert of Fig~\ref{fig:dynamic}a. This interaction geometry is chosen because the SHG of the probe is absorbed in the crystal and thus only the SHG generated at the back surface leaves the crystal. Due to the large attenuation at 2.1\,eV compared to the pump wavelengths chosen, the probed volume is uniformly excited.  

\section{Results}
\begin{figure}
\centering
\includegraphics{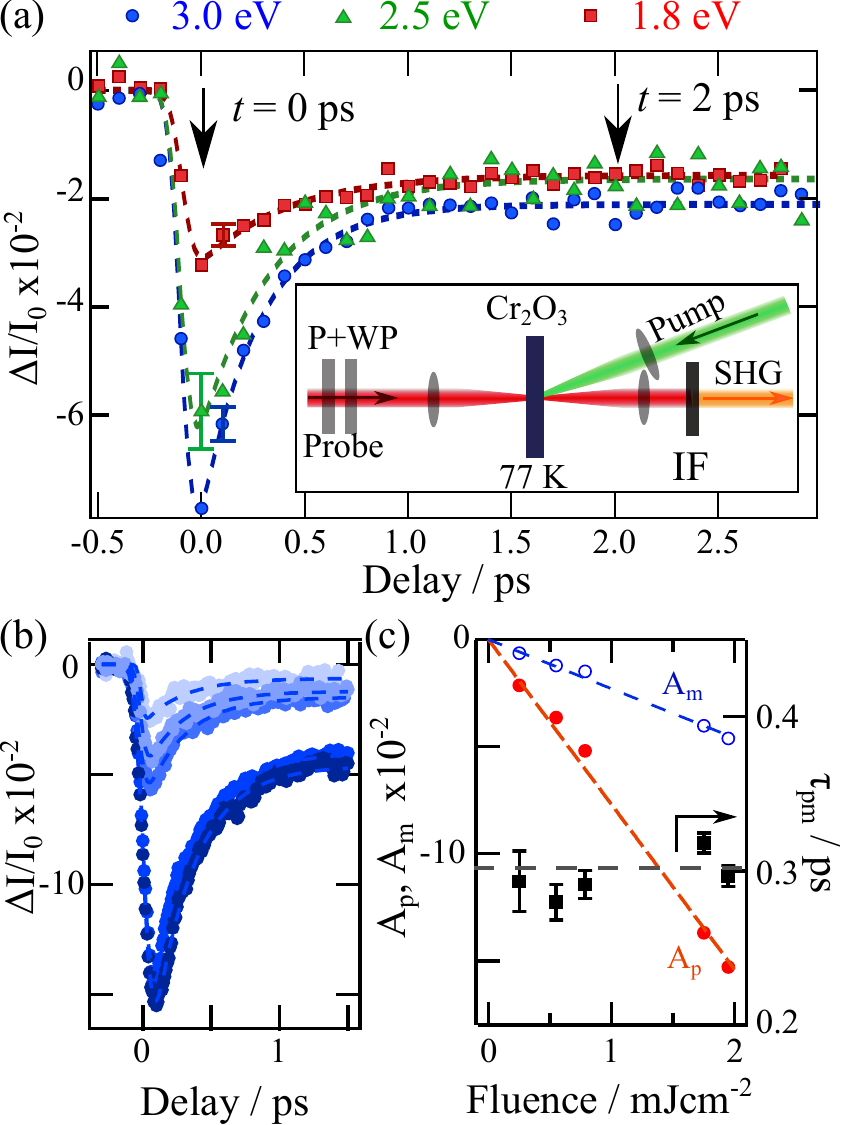}
\caption{(a) Visible-pump SHG-probe signal at the three different pump photon energies, 3.0\,eV (blue circles), 2.5\,eV (green triangles) and 1.8\,eV (red squares). Dashed lines correspond to the fits to Eq.~\ref{eq:fit}. Arrows indicate the times at which the polarization dependence shown in Fig.~\ref{fig:pol} were performed. The insert shows a schematic of the counter-propagating pump-probe setup (P - Polarizer, WP - Wave Plate, IF - Interference filters). (b) Fluence dependence of the SHG transient signal at 3\,eV from 0.25-2 mJ\,cm$^{-2}$ showing linear behavior. Dashed lines correspond to fits using Eq.~\ref{eq:fit}. (c) Fit parameters of the data in (b) showing that both the amplitude of the peak and plateau signal vary linearly with fluence. The $\tau_\mathrm{pm}$ time-constant is independent of fluence. \label{fig:dynamic}}
\end{figure}

Figure~\ref{fig:dynamic} shows the time-resolved change in the SHG signal ${\Delta I}/{I_0}$ for the three pump photon energies. In the traces shown, the pump, probe and SHG polarizations were all parallel to the crystallographic $x$-axis, however, other combinations were also measured. For all pump photon energies, a decrease in the SHG signal is measured. At delays $> 1$\,ps, all three signals settle to a plateau that lasts for more than 400\,ps. However, at shorter delays ($< 1$\,ps) additional fast dynamics are observed, which strongly depend on the pump photon energy. Due to the differences in the absorption coefficient for each pump pulse, the pump fluences were controlled in order to give roughly the same decrease in the SHG signal at 2\,ps. However, in all cases the transient signal was found to vary linearly with the pump fluence for all time delays as demonstrated in Figure~\ref{fig:dynamic}b for the case of 3\,eV excitation. Thus the different dynamics observed at early delays are not due to differences in the absorption and the response measured at 1.8\,eV excitation cannot be recreated by changing the intensity of a 3\,eV pump. 

The dynamics can be fit well by 
\begin{equation}
	\Delta I(t)/I_0 = \Theta(t)[A_\mathrm{p} e^{-t/\tau_\mathrm{pm}} + A_m(1-e^{-t/\tau_\mathrm{pm}})],
\label{eq:fit}
\end{equation}
where $\Theta(t)$ is the error function with a 75\,fs rise, fixed for all wavelengths, $A_\mathrm{p}$ and $A_\mathrm{m}$ are the magnitudes of changes at the spike at time zero and plateau respectively, and $\tau_\mathrm{pm}$ is the rapid recovery time constant, which was found to vary from $300\pm 20$\,fs at 3.0\,eV pumping to $400\pm 50$\,fs at 1.8\,eV pumping.  Figure~\ref{fig:dynamic}c shows that the time constant obtained and the ratio between the peak and plateau changes is independent of fluence, again demonstrating the linearity of the signal in this regime. We note that similar transients were observed when measuring the probe transmission at the fundamental photon energy (data not shown). However, we can exclude pump-induced changes in the fundamental intensity as the origin of the second harmonic dynamics because this would give rise to a signal that scales with the square of the pump fluence. As a result, transient changes in the linear optical properties have a negligible effect on the SHG transients.

\begin{figure}
\centering
\includegraphics{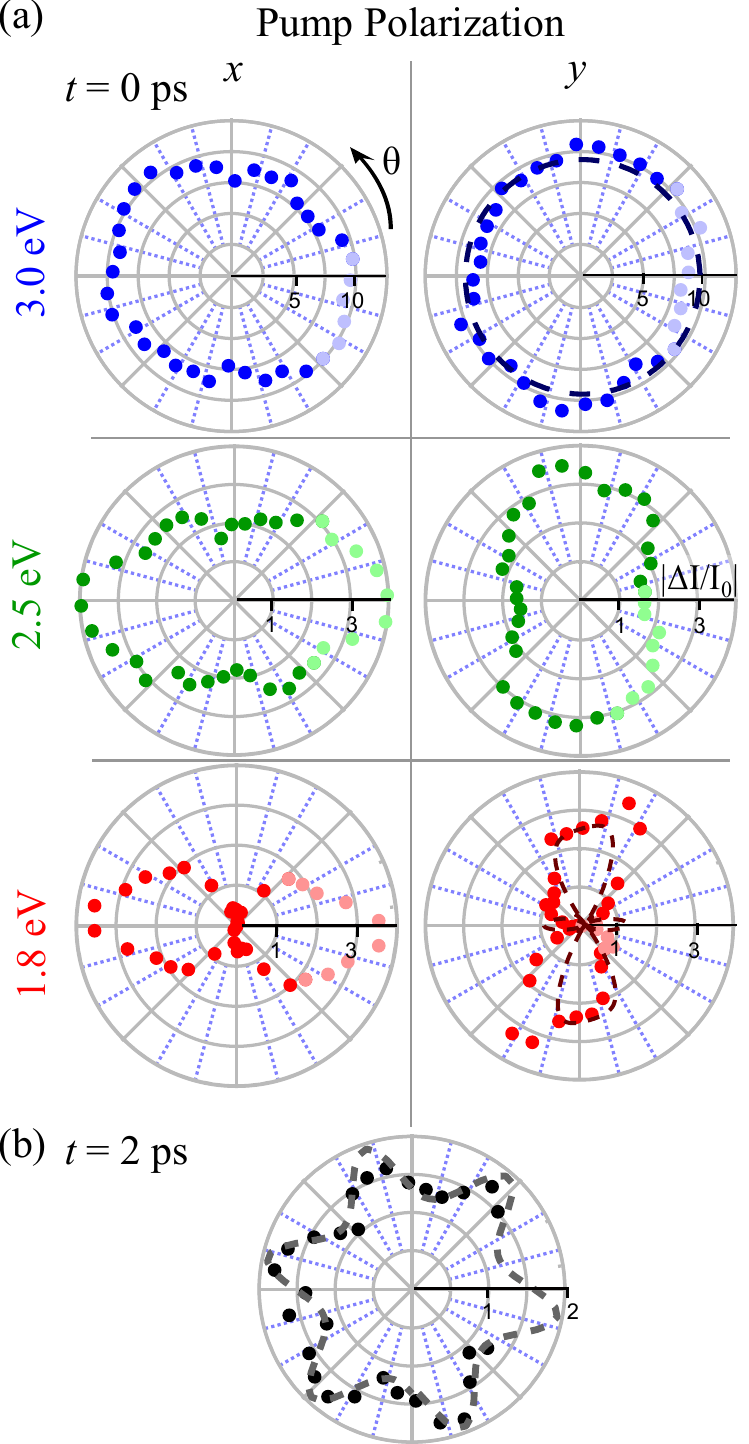}
\caption{(a) Change in the polarization dependence of the SHG probe at 0\,ps for two perpendicular pump polarizations ($x$, $y$). Lighter points are not measured and reconstructed from the symmetry of the signal. Dashed lines in the 3.0\,eV ($A_{1g}$) and 1.8\,eV ($E_g$) plots are fits for the $c_i$ coefficients at constant magnetization. (b) Change in the probe polarization dependence at 2 ps and corresponding fit to the scattering pattern when the AFM order parameter, $l$, is reduced by 0.75\%. \label{fig:pol}}
\end{figure}

In equilibrium, the intensity of the SHG signal can be directly related to the magnetic order. However, out of equilibrium this does not have to be the case~\cite{Huber2015}. In order to assign an origin to the observed dynamics we measured the polarization dependence of the transient SHG signal at the peak of the change and at long times after pumping, as shown in Fig.~\ref{fig:pol}.

When the system is in the long lived state, the SHG signal shows the same angular dependence irrespective of the pump photon energy or polarization. In this case the data can be accurately fitted by {\em only} reducing the antiferromagnetic order parameter, $l$, in Eq.~\ref{eq:angle}. Fig. ~\ref{fig:pol}b demonstrates the quality of the fit for $\Delta l = 0.75$\,\%. Thus the long-lived quenching of the SHG signal results from demagnetization. However, the angular dependence at short delays (Fig. ~\ref{fig:pol}a) shows a different behavior that strongly depends on the pump photon energy and polarization. This pattern cannot be fitted by changing $\Delta l$. Excitation at 3\,eV produces an isotropic change, while lowering the photon energy to 1.8\,eV breaks the symmetry of the SHG signal. Furthermore, the direction of the asymmetry can be controlled by the polarization of the pump. In this case it is clear that the SHG dynamics near time zero are not due to the spin system alone. 

\section{Discussion}
\begin{figure}
\centering
\includegraphics{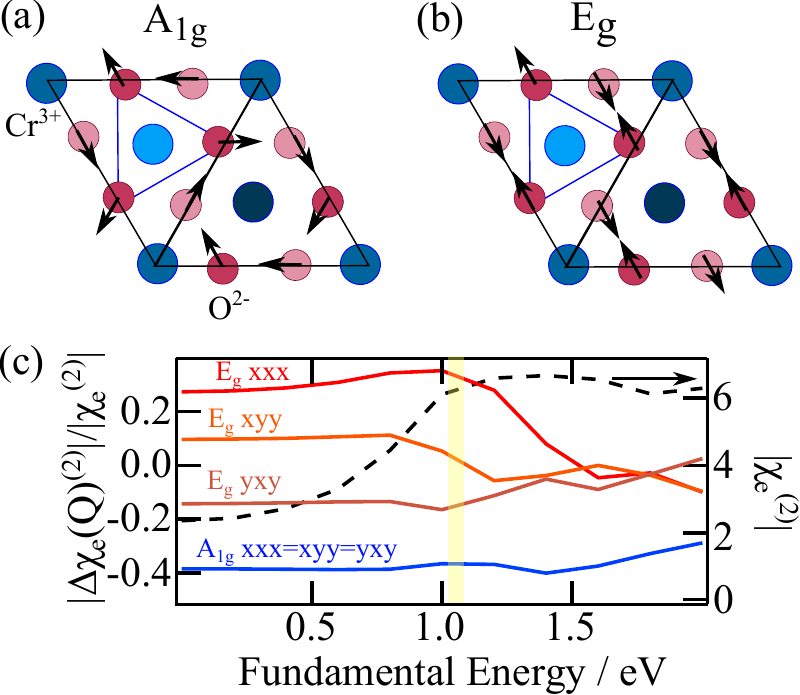}
\caption{(a) Sketch of the Cr$_2$O$_3$ unit cell in the $xy$-plane showing the $A_{1g}$ and (b) $E_g$  displacements of the oxygen ions around a Cr ion (darker colors correspond to atoms at different $z$ positions). (c) DFT calculations for changes in the magnitude of the nonlinear susceptibility for $A_{1g}$ and $E_g$ displacements of {0.05\,\AA}. The shaded area corresponds to the measured region. The dashed line shows the calculated equilibrium spectrum.\label{fig:theory}}
\end{figure}

As the excited states that are pumped are broadened due to electron-phonon coupling, we consider how crystal distortions can change the SHG signal, both in terms of amplitude and symmetry. Cr$_2$O$_3$ possesses $A_{1g}$ and $E_g$ Raman active modes, the motions of which are shown in  Figs.~\ref{fig:theory}a and~\ref{fig:theory}b. $A_{1g}$ modes preserve the symmetry of the crystal and thus can only modify the efficiency of the SHG process through a crystal-field-induced shift in the resonance condition and should not change the polarization dependence. For small displacements the $c_i$ coefficients change as as $c_i\rightarrow c_i (1+\delta Q_{A_{1g}})$, where $Q_{A_{1g}}$ is the phonon amplitude and $\delta$ is the coupling constant. Such a change can fit the polarization dependence measured for excitation at 3\,eV as shown in Fig.~\ref{fig:pol}a when the magnetization is kept constant and $\delta Q_{A_{1g}} = -0.05$.

$E_g$ displacements lower the crystallographic point group symmetry from $\bar{3}m$ to $2/m$, thus they can also change the polarization dependence as well as efficiency. As the $E_g$ mode is doubly degenerate, the direction along which the distortion occurs can be varied in the plane. Figure~\ref{fig:theory}b depicts the case for oxygen displacements along the $y$-direction. In this case the three non-zero components of the susceptibility become independent, which can be captured by $c_i^{jkl}\rightarrow c_i (1+\delta^{jkl}Q_{E_g})$, where $jkl= xxx,xyy,yxy$ are the indexes of the tensor elements. Thus a minimum of three parameters are needed to describe the effect.  The resulting change can fit the 1.8\,eV experimental data, with $c_i^{xxx}=0.9881c_i ~(\delta^{xxx}Q_{E_g}=-0.0119), c_i^{xyy}=1.0005c_i ~(\delta^{xyy}Q_{E_g}=0.0005)$ and $c_i^{xyx}=0.9915\,c_i ~ (\delta^{xyx}Q_{E_g}=-0.0085)$  as shown in Fig.~\ref{fig:pol}a. The perpendicular scattering pattern can be achieved when the symmetry breaking axis is rotated by 90$^\circ$. In this case the original non-zero components of the susceptibility remain unchanged and the corresponding components along the $y$ axis become non-zero. The data for pumping at 2.5\,eV can be described by a combination of the $A_{1g}$ and $E_g$ responses. 

These symmetry considerations are confirmed by calculations of the SHG signal under the distorted crystal structures obtained from density functional theory (DFT). We obtained the electronic structure for the ground state of Cr$_2$O$_3$ using the Abinit code, based on plane waves and pseudopotentials~\cite{Gonze2005,Gonze2009}. The antiferromagnetic state was taken into account in the calculations. Then the second order susceptibility was computed in Time-Dependent DFT using the 2light code based on the formalism developed in Ref.~\cite{Luppi2010}. The second order susceptibility is split into two parts, associated with the spin-up and spin-down components. Spin-orbit coupling is not included, as it is far beyond the reach of nonlinear ab initio calculations. Therefore, as expected, see for instance Ref.~\cite{Muto1998}, both components cancel. However, trends can be obtained for the effects of the crystal distortion by looking at one component only. The second order spin-up susceptibility for the distorted structures was obtained by moving the oxygen atoms along the $A_{1g}$ and $E_g$ modes at the level of the DFT calculations. The spectra were calculated in the independent particle approximation. Convergence was achieved with 432 off-symmetry shifted $k$ points in the full Brillouin zone and 90 unoccupied states. 

The results are shown in Fig.~\ref{fig:theory}c. The $A_{1g}$ displacement induces an equal decrease of all elements, whereas the $E_g$ distortion breaks the degeneracy of the elements, as expected from the above symmetry analysis. Furthermore, these calculations enable us to estimate the size of the induced displacements from the magnitude of our signal. The calculations were performed with a displacement of {0.05\,\AA} and induce changes that are larger than those measured experimentally, thus resonant SHG is extremely sensitive to highly symmetric atomic motion.

\begin{figure}
\centering
\includegraphics{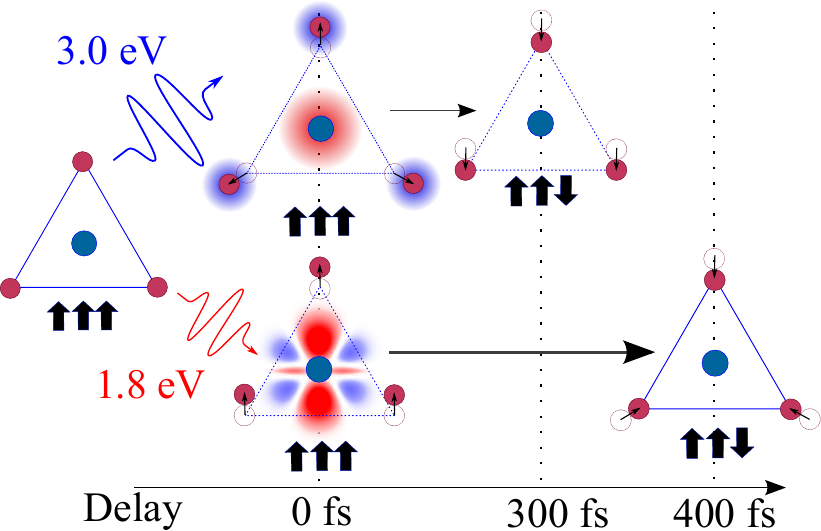}
\caption{Part of the unit cell centered on the Cr ion and nearest oxygen neighbors. Bold arrows correspond to the spin state.  Photoexcitation at 1.8\,eV excites electrons from the $t_{2g}$ levels into the $e_g$ levels while preserving the spin state. The resulting charge redistribution is anisotropic (blue corresponds to more positively charged regions, red more negative) and couples strongly to the anisotropic $E_g$ phonon mode. The electrons are then scattered to a spin-flipped $t_{2g}$ state in 400\,fs. Thus the distortion is relaxed and the system is demagnetized. Excitation at 3\,eV causes charge transfer from O to Cr. This causes a symmetric charge redistribution mapping onto the $A_{1g}$ phonon mode and scattering occurring within 300\,fs. \label{fig:story}}
\end{figure}

These results enable us to build a picture for the demagnetization process, which is sketched in Fig.~\ref{fig:story}. The pump pulse excites electrons from the occupied $^4A_2$ ground state which is composed of the $t_{2g}$ levels of the Cr ion, into final states that depend on the photon energy. Lower energy photons transfer electrons into the unoccupied T states, which mix the $3d-e_g$ and $t_{2g}$ levels of the Cr ion~\cite{Muto1998}, changing the charge distribution in the vicinity of the Cr ion. As these orbitals are anisotropic, neighboring ions experience a different force, causing oxygen atomic motion that is along the $E_g$ phonon coordinate. Higher energy photons trigger transitions that are more of a charge transfer character, i.e. charge is redistributed between chrome and oxygen ions. In the limit that the charge transfer is complete, this produces a symmetric force on the oxygen ions which couples to $A_{1g}$ motion. The preferential coupling to the $A_{1g}$ mode for high photon energies is also confirmed by resonant Raman scattering measurements shown in Fig~\ref{fig:Static}d. Furthermore, the timescale for this processes is in good agreement for the equivalent process that occurs in ruby~\cite{Fonger1975}.

As both Raman modes have frequencies in the region 9 - 18\,THz, they respond adiabatically during the excitation pulse and coherent motion of the lattice is not triggered. In the distorted state, the excited electrons in both cases can be efficiently scattered into the spin-flipped $^2E$ states of Fig.~\ref{fig:Static}a. These states have the same spatial distribution of charge as the ground state, but the electron's spin is flipped~\cite{Muto1998}. Therefore, the force causing the displacements is lost and the crystal relaxes, while at the same time the magnetization is reduced. Once relaxed, the equilibrium spin-lattice relaxation process dominates, which is  much slower. These results differ from previous measurements of spin dynamics in Cr$_2$O$_3$~\cite{Satoh2007, Dodge1999}, since we operate at lower temperatures and lower excitation fluences. Higher fluences and temperatures may enable different scattering processes. However, the 300-400\,fs timescales obtained for Cr$_2$O$_3$ are very similar to the timescale observed in antiferromagnetic order melting in insulating Sr$_2$IrO$_4$ measured using time-resolved resonant diffraction~\cite{Dean2016}, thus demonstrating that rapid control of the magnetic state is possible in antiferromagnetic insulators when exciting above the band gap. 

\section{conclusion}
In summary, we have shown how specific lattice modes dictate the demagnetization rate in Cr$_2$O$_3$ and that the lattice modes can be controlled by resonantly tuning the light pulse in the visible spectral region, enabling a modulation of the demagnetization rate by 25\%. This control is achieved through the presence of in-gap states that occur within the band gap of insulating materials. Through the control of the crystallographic structure, presented here, or through direct excitation of the magnetic subsystem with below gap excitation~\cite{Bossini2014}, insulating antiferromagnets present new opportunities for light control that are not available in ferromagnetic metals. As the coupling between charge excitations and Raman active modes is direct and occurs in the optical regime, it also has the potential to be more efficient and easier to implement that selective methods that require non-linear coupling between IR and Raman active modes~\cite{Forst2011a}.

Furthermore, we have shown that, out of equilibrium, time resolved SHG is a powerful tool for monitoring {\em both} magnetic and structural degrees of freedom and is capable of measuring small, symmetric atomic displacements. When combined with theory, the ability to observe small changes in the lattice simultaneously with the evolution of the magnetic degree of freedom, enables measurements in a standard laboratory which would otherwise require large scale user facilities.

\begin{acknowledgments}
 The research leading to these results has received funding from LASERLAB-EUROPE (grant agreement no. 284464, EC's Seventh Framework Program). We acknowledge GENCI (project 544) for the computational support provided. SW received financial support from Ramon y Cajal program RYC-2013-14838 and Marie Curie Career Integration Grant PCIG12-GA-2013-618487. VGS, TAM and SW acknowledge support from Severo Ochoa Excellence Grant and Fundaci\`{o} Privada Cellex. GC acknowledges support by the European Union’s Seventh Framework Program (FP7/2007-2013) Grant No. CNECT-ICT-604391 (Graphene Flagship).
\end{acknowledgments}

\end{document}